\documentclass{JHEP3}
\usepackage{amsmath}
\usepackage{axodraw}
\usepackage{epsfig}
\usepackage{amssymb}
\usepackage{mathrsfs}
\usepackage{mathcomp}
\usepackage{cite}
\usepackage{rotating}
\usepackage[latin1]{inputenc}

\newcommand{\abar}{\bar{\alpha}}
\newcommand{\as}{\alpha_s}

\preprint{
}

\title{On the High Energy Behaviour of The Total Cross Section 
  in the QCD Dipole Model}

\author{Emil Avsar\\
  Institut de Physique Th\'eorique de Saclay,
  F-91191 Gif-sur-Yvette, France \\
  E-mail: \email{Emil.Avsar@cea.fr}}

\abstract{ 
In this paper we perform a numerical study of the tranverse expansion of
hadronic scattering amplitudes in the dipole picture of high energy QCD. We go beyond
the mean field approximation by including fluctuations and also wave function saturation
effects, and the evolution with both a fixed and a running coupling is investigated. We also
study the nonperturbative aspects, and as has been predicted earlier, our results indicate
that the Froissart-Martin bound is saturated once confinement effects are included in the
evolution. Thus the total cross section increases proportional to the square of the logarithm
of the cms energy. Using our proton model developed earlier we furthermore see that we
obtain a reasonable value for the proportionality coefficient. The impact of saturation and
non-leading effects on this coefficient is also studied.}

\begin{document}

\sloppy

\section{Introduction}
\label{sec:intro}

The Froissart-Martin (FM) theorem states that hadronic total cross 
sections must satisfy 
\begin{eqnarray}
\sigma_{tot}(s) \leq C\cdot \mathrm{ln}^2 (s/s_0) \,\,\, \mathrm{as} 
\,\,\, s\to \infty,
\label{eq:FM}
\end{eqnarray}
where the coefficient $C$ can be estimated as $C \sim 1/m_\pi^2$.
The proof of this theorem relies on some general properties such 
as the unitarity of the $S$-matrix, the existence of a mass gap, 
and the possibility of using subtracted dispersion relations. Although 
not strictly proven within QCD, it is widely believed that 
this bound should indeed be true for the strong interactions.

The high energy evolution equations (the Balitsky-JIMWLK, or B-JIMWLK hierarchy) 
for the hadronic amplitudes $T_s(\pmb{b})$ were derived in \cite{Balitsky:1995ub}, 
and these equations can also be described by the 
Color Glass Condensate (CGC) formalism \cite{Iancu:2002xk}, or, in 
the large $N_c$ limit, by the simpler dipole 
formalism \cite{Mueller:1993rr}. 
The solution to these equations exhibits saturation 
at each $\pmb{b}$, 
\emph{i.e.} $T_s(\pmb{b}) \leq 1$, with  $T_s(\pmb{b}) = 1$ 
being the black disc limit. This condition is, however, not 
sufficient in order to satisfy the FM bound, since 
the FM bound is relevant for the total cross section 
which involves an integration over all $\pmb{b}$. In fact, 
it is quite obvious that if one naively takes the solution 
to the perturbative evolution equations for $T_s(\pmb{b})$, 
and then integrates over all $\pmb{b}$ to get the 
total cross section as 
\begin{equation}
\sigma_{tot}(s)=2\,\int d^2\pmb{b}\, T_s(\pmb{b}),
\end{equation}
one will most certainly violate the FM bound, 
since the interaction is mediated by massless
gluons with a Coulomb like behaviour even at large distances.	
Indeed for an interaction mediated by a massless 
particle, such as the photon in QED, 
the coefficient $C$ in \eqref{eq:FM} is infinite, since 
$m_\gamma=0$.  

In this paper we will study the behaviour of hadronic 
cross sections with respect to the FM bound. We will 
use our model developed in \cite{Avsar:2005iz, Avsar:2006jy, Avsar:2007xg}, 
which is based on the 
QCD dipole model, to calculate the growth of the total $pp$ 
cross section. In \cite{Avsar:2006jy} we have suggested the dipole swing
mechanism in order to take into account the missing 
saturation effects in the dipole cascade evolution, 
and results show that we obtain an almost frame independent
evolution. As we will later discuss in section \ref{sec:swing}, the 
dipole swing has some similarities with the saturation 
mechanism in the CGC formalism. Furthermore, the swing
affects the expansion of the dipole cascade in the 
transverse plane, 
and it is therefore interesting to see how 
large effects it has on the transverse expansion of 
the scattering amplitude.  

Our main results are presented in section 
\ref{sec:res}. We will see that the confinement mechanism is 
crucial in order to obtain sensible results. This 
is especially the case when a running coupling is used. Obviously 
this is to be expected, as otherwise the cascade evolution 
favours the formation of too large dipoles. Without 
confinement the 
cross section grows exponentially in $Y$ (defined as $Y = \mathrm{ln}(s/s_0)$, 
with $s_0 \approx $1 GeV$^2$), as expected from the 
long ranged nature of the massless gluon fields. 

The leading order cascade evolution is 
strongly suppressed by the non-leading 
effects \cite{Avsar:2007xg}. Besides the running 
coupling, the non-leading effects come from the non-singular 
terms in the gluon splitting function $P(z)$, and 
the so-called energy scale terms which are related to the 
conservation of $p_+$ and $p_-$ respectively \cite{Avsar:2007xg} 
(exact energy-momentum conservation goes beyond the 
NLO corrections, however).
Once these effects are included (using the prescription 
described in \cite{Avsar:2005iz}), we see that 
 the growth of the cross section is much 
reduced. Nevertheless,
when increasing the energy, one can see that the growth 
is still faster than what is permitted by the FM bound. 
Interestingly, we will in this case see that $\sigma_{tot}$ can be 
fitted rather well by a polynomial in $Y$, for $\alpha_s = 0.2$
and $Y$ up to around 32 units.  

As the running coupling is a non-leading effect, 
it might seem strange that the growth of the cross section 
is much faster as compared to the fixed 
coupling case. This is mainly due to two reasons. First, 
the value of $\alpha_s$ 
gets very large during the evolution (especially in the 
specific model which we use), signaling the breakdown
of the perturbative approach. Secondly, the total growth is 
faster due to unrealistically large 
contributions from larger $b$. However, the growth of $T_s(b=0)$ is 
actually slower in the running coupling case.

The transverse expansion of the dipole cascade is of course not really 
consistent with QCD at distances larger than the confinement scale, 
since it is driven by Coulomb fields. 
We will therefore also study the expansion when confinement effects 
are modeled by replacing the Coulomb propagators, $1/\pmb{k}^2$, 
with screened propagators, $1/(\pmb{k}^2+M^2)$. In this 
case a ln$^2s$ growth is obtained, and the FM bound is thus saturated.
We also see that 
we get a very sensible result for the coefficient $C$ in 
\eqref{eq:FM}. The fact that $C$ can be estimated by 
combining the perturbative growth with nonperturbative 
initial conditions has been proposed in \cite{Ferreiro:2002kv}, although 
the leading order BFKL result gives a way too large 
value for $C$. 



\section{The growth of the Black Disc}

Let us consider a hadronic projectile impinging on 
a hadronic target. The projectile might here be an
elementary colour dipole, and the target a proton or
a nucleus. We will from
now on denote the generic scattering amplitude between an arbitrary projectile and an
arbitrary target by $T_Y(\pmb{b})$. When the initial (at $Y=0$) 
projectile or target is fixed we also
use additional parameters to denote the amplitude. For example if the initial projectile is a
dipole of size $r$ which scatters off an arbitrary target, then we write
$T_Y(\pmb{r}, \pmb{b})$. If we have the
scattering of two dipoles $r$ and $r_0$ we instead write 
$T_Y(\pmb{r}, \pmb{r}_0, \pmb{b})$.
The region in $\pmb{b}$ where the scattering 
amplitude $T_Y(\pmb{b})$ satisfies $T_Y(\pmb{b}) \approx 1$ 
is called the black disc region, since in this case
the projectile is strongly absorbed by the target. The region 
where $T_Y(\pmb{b}) \approx 0$ is on the other hand referred 
to as the ``white'' region\footnote{The white region can be defined 
as the region where the scattering between a dipole of arbitrary size
and the target is small, \emph{i.e.} 
where the local saturation scale $Q_s(\pmb{b})$ 
is smaller than, or the order of, 
$\Lambda_{QCD}$ \cite{Ferreiro:2002kv}.} since here the target appears transparent. 
In the region between, the scattering 
is ``grey''. Specifically, the black disc region is defined 
as the radius of the disc\footnote{The average amplitude 
is isotropic in the transverse plane. On an event-by-event 
basis, however, there is no isotropy.} within which the average amplitude satisfies 
\begin{equation}
T_Y(\pmb{b}) \geq a \, \, \, \mathrm{for}\,\,\, 
|\pmb{b}| \leq R_{bd}(Y), \,\,\,\, 
\mathrm{where} \, \, \, \, a \approx 0.5.
\label{eq:blackdiscdef}
\end{equation}
The total cross section can then be estimated as
\begin{eqnarray}
\sigma_{tot}(Y) = 2\, \int d^2\pmb{b} \, T_Y(\pmb{b}) 
\sim 2\pi R^2_{bd}(Y),
\end{eqnarray}
where $R_{bd}(Y)$ is the radius of the black disc at 
rapidity $Y$. (Obviously $R_{bd}(Y)$ is also dependent 
on the projectile and the target, but we do not explicitely write this 
dependence.) If $\sigma_{tot}$ is to 
satisfy the FM bound, it is seen that $R_{bd}(Y)$ can 
at most grow linearly with $Y=$ ln$(s/s_0)$. In what follows, 
we will study the $Y$ dependence for $R_{bd}(Y)$ for different 
situations, using our model developed in 
\cite{Avsar:2005iz, Avsar:2006jy, Avsar:2007xg}.

\subsection{The BFKL Growth}
\label{sec:bfkl}

The $Y$ dependence of $R_{bd}(Y)$ which follows from the 
solution to the non-linear QCD evolution equations have 
been discussed in \cite{Ferreiro:2002kv, Kovner:2002yt, Kovner:2002xa, 
Ikeda:2004zp}, in the context of the Balitsky--Kovchegov 
(BK) equation \cite{Balitsky:1995ub, Kovchegov:1999yj} which is a mean field version 
of the B-JIMWLK hierarchy. A detalied
numerical study of the BK equation with impact parameter dependence was performed
in \cite{GolecBiernat:2003ym}. 
These works have demonstrated that the BK equation leads to an exponential growth of
$R_{bd}(Y)$, even if one starts with an initial profile in $\pmb{b}$
 which falls off very steeply. Using the dipole language, the fast growth of 
$R_{bd}(Y)$ can be understood as follows. Assume that, 
for a given dipole projectile of size $r$,
one studies the evolution for $b \equiv |\pmb{b}| >> R_{bd}(Y)$, where 
$T_Y(\pmb{r},\pmb{b}) << 1$. The BK equation can then be replaced 
by the linear BFKL equation\footnote{Strictly speaking 
one has to be careful when linearizing the BK equation since 
there might be contributions 
to the evolution from inside the black region where 
the scattering is strong, for details see \cite{Ferreiro:2002kv, Kovner:2002yt}.
} whose solution in $\pmb{b}$
can be written as
\begin{eqnarray}
T_Y(\pmb{r},\pmb{r}_0,\pmb{b}) = \int d^2\pmb{b}'\int \frac{d^2\pmb{r}'}{2\pi r'^2}
n_Y(\pmb{r}',\pmb{b}', \pmb{r}_0)\, T_0(\pmb{r},\pmb{b}| \pmb{r}',\pmb{b}'),
\end{eqnarray}
where $n_Y$ is the dipole density of the target at $Y$, 
and $T_0$ is the basic dipole--dipole scattering 
amplitude. 
Here we have assumed that the target initially consists 
of a single dipole of size $r_0$. Of course one could 
imagine more complicated initial conditions for the 
target, but as long as the scattering is weak, the 
precise choice should not matter.
The elementary dipole--dipole scattering amplitude, $T_0$, can be simplified when the 
separation of the dipoles are large compared to their 
sizes (remember that we assume $b >> R_{bd}$), 
in which case it decays as $1/|\pmb{b}-\pmb{b}'|^4$.
It can then be shown that 
\begin{eqnarray}
T_Y(\pmb{r}, \pmb{r}_0,\pmb{b}) &\sim& \alpha_s^2r^2n_Y(r, b, r_0) \nonumber \\
 &\sim & 32\alpha_s^2\,\frac{\mathrm{log}\frac{16b^2}
{r_0r}}{(\pi c^2Y)^{3/2}}\,\,\mathrm{exp}\biggr ( 
\omega Y - \mathrm{log}\frac{16 b^2}{r_0r}
- \frac{\mathrm{log}^2\frac{16 b^2}{r_0r}}{c^2Y}
\biggl ).
\label{eq:Tbfkl}
\end{eqnarray}
Here $\omega = 4$ln2$\cdot \alpha_s$, $c^2= 14\zeta(3)\bar{\alpha}$, and 
$\bar{\alpha}\equiv \alpha_sN_c/\pi$.
Omitting constants and the slowly varying prefactors, we can write 
\begin{eqnarray}
T_Y(\pmb{r}, \pmb{r}_0, \pmb{b}) \sim \frac{r_0r}{b^2}\,\mathrm{exp}\biggr (
\omega Y - \frac{\mathrm{log}^2\frac{16 b^2}{r_0r}}{c^2Y}
\biggl ).
\end{eqnarray}
Using definition \eqref{eq:blackdiscdef}, 
it can now easily be seen that this formula implies an 
exponential growth for $R_{bd}(Y)$, simply 
because the power-like decay in $b$, coming 
from the Coulomb fields associated with the exchanged 
gluons, is too slow to compensate for the fast growth in $Y$.  

\subsection{Saturation Effects in The Cascade Evolution}
\label{sec:swing}
\subsubsection{Multiple Scatterings and Boost Invariance}

The discussion so far has neglected saturation effects 
in the dipole cascade evolution. In the dipole model, 
unitarization at each $\pmb{b}$ is obtained by taking into
account multiple dipole interactions. In an eikonal 
approximation the multiple scatterings can be summed to all
orders, with the result that $T$ can be written as 
\begin{eqnarray}
T_Y(\pmb{b}) = \biggl \langle 1 - \mathrm{exp} \biggl (
-\sum_i \sum_j T_0(\pmb{x}_i, \pmb{y}_i| \pmb{u}_i, \pmb{v}_i )
\biggr ) \biggr\rangle.
\end{eqnarray}
Here the brackets denote averaging over different events (notice that
it is the event by event amplitude which exponentiates). The sums 
over $i$ and $j$ just denote sums over the dipoles in the individual 
dipole cascades, and the $\pmb{b}$ and $Y$ dependences are 
implicit in the right hand 
side. Thus we explicitely have $T_Y(\pmb{b}) \leq 1$ at each $\pmb{b}$. 
This condition is absolutely necessary for our discussion, as 
otherwise it would make no sense to talk about the black disc limit
or the Froissart bound. The Froissart bound determines how rapidly 
the black disc can expand in $\pmb{b}$, but if $T$ is not bounded 
by 1 then there is no black disc limit at all. 

Even though one can unitarize $T$ at each $\pmb{b}$, the evolution of 
the individual dipole cascades satisfy the 
linear BFKL equation. Since what appears as multiple scatterings in 
one frame will appear as saturation effects in the cascade evolution 
(\emph{i.e.} in the onium wavefunction) 
in another frame, the formalism is not frame independent. The discussion above should 
therefore be generalized to the case including saturation
effects also in the cascade evolutions. Our 
results presented below includes such effects, and we therefore
discuss them in this section. 

\subsubsection{Dipole Swing}

We have previously argued that one can take into account saturation effects in the dipole
cascade evolution by including the so-called dipole swing into the formalism. We will here
once again discuss the idea behind the swing. An analytical proof for boost invariance is,
however, still lacking. We have therefore so far implemented an approximation in our Monte
Carlo (MC) code (as have been discussed in \cite{Avsar:2006jy, Avsar:2007xg}), 
and we will at the end of this section
try to sketch the similarities between our implementation and the saturation mechanism
in the CGC formalism.

The generic evolution equations for high energy QCD 
including all possible pomeron interactions are 
not yet known. Equations have, however, been derived 
for simple toy models which neglect the complicated 
topology of the full model \cite{Blaizot:2006wp, 
Iancu:2006jw}. 
The saturation mechanism in these simpler models 
have similarities with the CGC formalism, and they 
can also be formulated in a stochastic evolution 
with similarities to the dipole evolution in the 
presence of the dipole swing, as 
we now explain. 

These equations can namely be interpreted in terms of (positive 
definite) $k\to k+1$ \linebreak vertices \cite{Avsar:2007xh}. The dipole swing (discussed
in more detail below) is 
a process which instantaneously in $Y$ replaces two initial dipoles,
$(x_1,y_1)$ and $(x_2,y_2)$,  
with two final dipoles, $(x_1,y_2)$ and $(x_2,y_1)$. 
Here ($y$)$x$ denotes the transverse position of the 
(anti-)colour end of the dipole (here we do not use boldface letters for the vectors).
The generic $k\to k+1$ vertices can then be constructed by 
combining the usual $1\to 2$ dipole splitting with $k-1$ 
instantaneous dipole swings. In the toy models in \cite{Blaizot:2006wp, 
Iancu:2006jw} these
vertices give a boost invariant evolution. However, the amount of information we can extract from
the toy models is limited. In QCD it is important to take into account the colour degrees
of freedom which are completely absent in the toy models. The colour structures of the
multiple scattering diagrams were discussed in detail in \cite{Avsar:2007xh} 
and we will here again briefly discuss the colour structures.

The linear cascade evolution is directly related to the leading $N_c$ approximation. The
dipole splitting kernel is proportional to $\as N_c = \abar$ (for simplicity we here neglect the
factor $\pi$ in the definition of $\abar$ as it is completely irrelevant for our discussion), while the
scattering diagrams are proportional to $\as^2 = \abar^2/N_c^2$. To take into account the effects of
multiple scatterings in all frames one would therefore need to include processes proportional
to $\abar \cdot \abar^{2n}/N_c^{2n}$ 
(the first factor of $\abar$ comes from the dipole splitting) in the cascade evolution.
In the leading $N_c$ approximation such processes are absent\footnote{
The reason one includes multiple scatterings is because they are dominant at high energies even if they
are colour suppressed. The contribution from $n$ pomeron exchange goes like $e^{n\omega Y}$ 
as compared to single
pomeron exchange which goes like $e^{\omega Y}$ , where $\omega$ is the BFKL intercept.}.  
To correctly include saturation effects
in the cascade evolution it is therefore very important that one studies the colour structures
of the relevant Feynman diagrams.

\FIGURE[t]{
\scalebox{1.1}{\mbox{
\begin{picture}(300,200)(0,0)

\Vertex(0,150){2}
\Vertex(100,150){2}
\CArc(15,150)(15,90,180)
\CArc(15,150)(15,180,270)
\CArc(85,150)(15,0,90)
\CArc(85,150)(15,270,0)
\Line(15,165)(85,165)
\Line(15,135)(85,135)

\Text(0,165)[]{$x_1$}
\Text(5,132)[]{$y_1$}
\Text(0,115)[]{$x_2$}
\Text(7,80)[]{$y_2$}
\Text(15,34)[]{$u$}
\Text(15,-5)[]{$v$}
\Text(20,146)[]{$z$}

\Text(200,165)[]{$x_1$}
\Text(205,132)[]{$y_1$}
\Text(200,115)[]{$x_2$}
\Text(207,80)[]{$y_2$}
\Text(215,34)[]{$u$}
\Text(215,-5)[]{$v$}
\Text(220,146)[]{$z$}

\GlueArc(30,165)(15,180,270){2}{4}
\Gluon(30,150)(70,150){2}{6}
\GlueArc(70,135)(15,0,90){2}{4}
\Gluon(25,135)(25,115){2}{3}
\Gluon(75,135)(75,85){2}{6}
\Gluon(38,135)(38,0){2}{17}
\Gluon(65,135)(65,0){2}{17}
\DashLine(51,170)(50,-5){2}

\Vertex(0,100){2}
\Vertex(100,100){2}
\CArc(15,100)(15,90,180)
\CArc(15,100)(15,180,270)
\CArc(85,100)(15,0,90)
\CArc(85,100)(15,270,0)
\Line(15,115)(85,115)
\Line(15,85)(85,85)

\Vertex(0,15){2}
\Vertex(100,15){2}
\CArc(15,15)(15,90,180)
\CArc(15,15)(15,180,270)
\CArc(85,15)(15,0,90)
\CArc(85,15)(15,270,0)
\Line(15,30)(85,30)
\Line(15,0)(85,0)

\Vertex(200,150){2}
\Vertex(300,150){2}
\CArc(215,150)(15,90,180)
\CArc(215,150)(15,180,270)
\CArc(285,150)(15,0,90)
\CArc(285,150)(15,270,0)
\Line(215,165)(285,165)
\Line(215,135)(285,135)

\GlueArc(230,165)(15,180,270){2}{4}
\Gluon(230,150)(270,150){2}{6}
\GlueArc(270,135)(15,0,90){2}{4}
\Gluon(225,135)(225,115){2}{3}
\Gluon(275,165)(275,85){2}{10}
\Gluon(238,135)(238,0){2}{17}
\Gluon(265,135)(265,0){2}{17}
\DashLine(251,170)(250,-5){2}

\Vertex(200,100){2}
\Vertex(300,100){2}
\CArc(215,100)(15,90,180)
\CArc(215,100)(15,180,270)
\CArc(285,100)(15,0,90)
\CArc(285,100)(15,270,0)
\Line(215,115)(285,115)
\Line(215,85)(285,85)

\Vertex(200,15){2}
\Vertex(300,15){2}
\CArc(215,15)(15,90,180)
\CArc(215,15)(15,180,270)
\CArc(285,15)(15,0,90)
\CArc(285,15)(15,270,0)
\Line(215,30)(285,30)
\Line(215,0)(285,0)

\end{picture}
}}
\caption
{\label{fig:swing1}  Two diagrams contributing to interactions within the same cascade as explained in the
text. The dashed line indicates the cut between the amplitude and the comple conjugate amplitude.} 
}

\FIGURE[t]{
  \includegraphics[angle=0, scale=0.58]{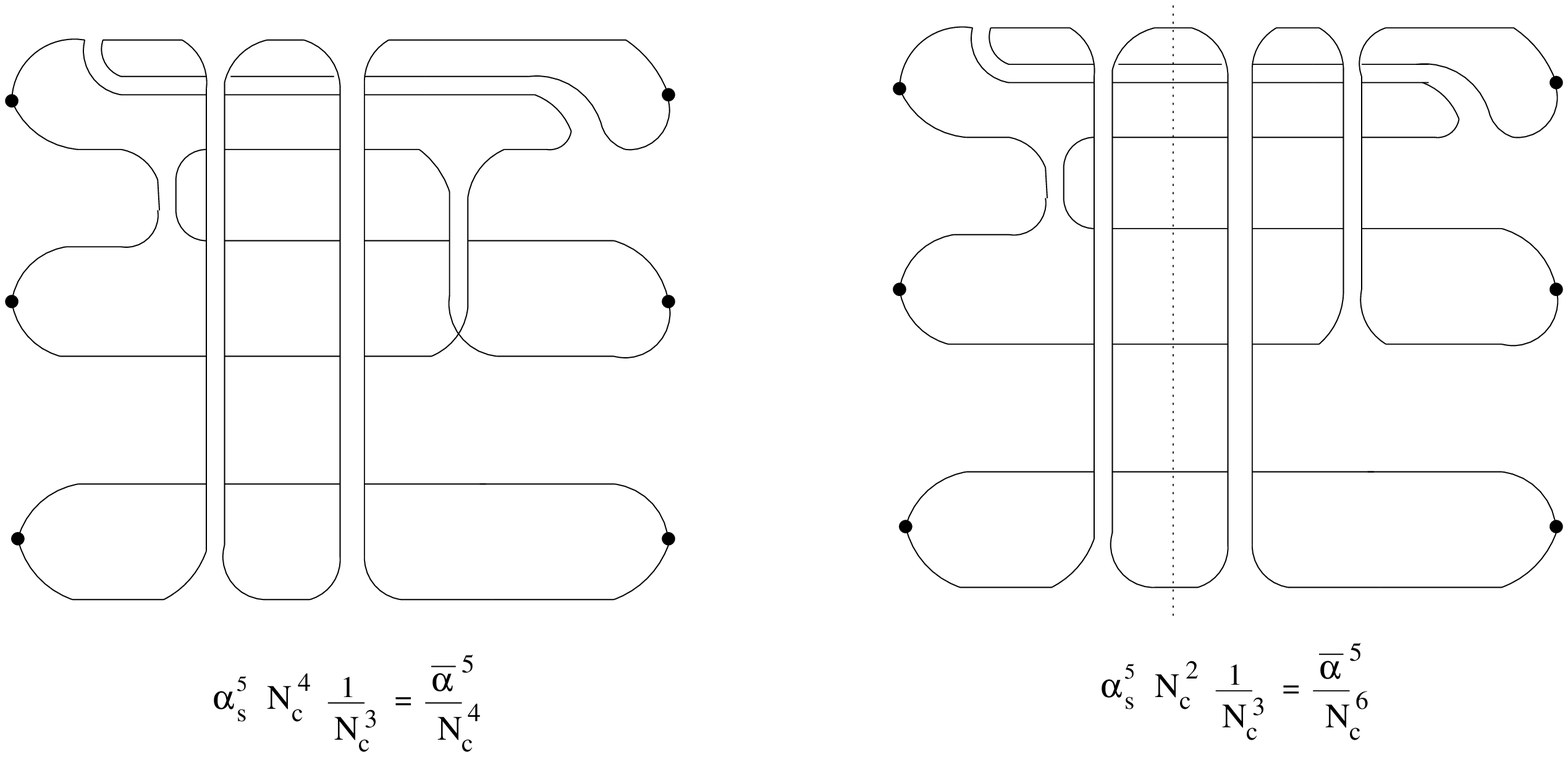}
\caption{\label{fig:swing2}  The respective colour structures of the two diagrams from figure 
\ref{fig:swing1}.}
}

Let us consider the situation in figure \ref{fig:swing1}. Here we have two right moving dipoles,
$(x_1, y_1)$ and $(x_2, y_2)$, and one left moving dipole $(u, v)$. When rapidity is increased by
$\Delta Y$, 
the gluon $z$ can be emitted from $(x_1, y_1)$ (thus we put the evolution into the right moving
system). In the left diagram in figure \ref{fig:swing1} we show just one of the contributions to this
process. Here $z$ is in the amplitude emitted by the quark located at $x_1$, and is absorbed
in the complex conjugate amplitude by the antiquark located at $y_1$. In addition to this we
have two-gluon exchange between the system consisting of $x_1, y_1$ and $z$, and $(x_2, y_2)$. 
In the amplitude a gluon is exchanged between the antiquark at $y_1$ and the quark at $x_2$, while
in the complex conjugate amplitude a gluon is exchanged between the antiquark at $y_1$ and
the antiquark at $y_2$. These processes contribute to the evolution of the wavefunction of
the right moving system. In addition, there is a two-gluon exchange between the right and
left moving systems. In the amplitude there is a gluon exchanged between the quark at
$x_1$ and the oppositely moving antiquark at $v$, and there is again a gluon exchange between
$x_1$ and $v$ in the complex conjugate amplitude. 
In the right diagram in figure \ref{fig:swing1}, the gluon
exchanged between $y_1$ and $y_2$ in the complex conjugate amplitude is instead exchanged
between $x_1$ and $y_2$.

In figure \ref{fig:swing2} we show the respective colour structures of the two diagrams from figure \ref{fig:swing1}.
Counting the vertices and the colour loops we see that the left diagram is proportional to
$\abar^5/N_c^4$.
The extra factor $1/N_c^3$ comes from the black dots where we project out the colour
singlet contributions (a dipole is a colour singlet). In the right diagram we can count 2
loops, and the process shown is therefore proportional to $\abar^5/N_c^6$ and thus suppressed as
compared to the left diagram.

These diagrams contribute to processess which can be interpreted as follows. The left
diagram in figure \ref{fig:swing1} is one of the diagrams which contribute to the process where $(x_1, y_1)$
first splits into $(x_1, z)$ and $(y_1, z)$ by the emission of the gluon $z$, after which the dipoles
$(z, y_1)$ and $(x_2, y_2)$ exchange a gluon, whereby they are replaced by two new dipoles, $(z, y_2)$
and $(x_2, y_1)$. Then finally the dipole $(x_1, z)$ interacts with the target $(u, v)$. This process
has therefore a dipolar interpretation, and the step where $(z, y_1)$ and $(x_2, y_2)$ are replaced
by $(z, y_2)$ and $(x_2, y_1)$ precisely describes the dipole swing. The right diagram in figure \ref{fig:swing1}
can on the other hand not be described in terms of dipoles. Here a gluon is exchanged
between $(z, y_1)$ and $(x_2, y_2)$ in the amplitude, but in the complex conjugate amplitude a
gluon is instead exchanged between $(x_1, z)$ and $(x_2, y_2)$. However, we also see that this 
process is suppressed, since it goes like $\abar^5/N_c^6$ instead of $\abar^5/N_c^4$.

In case the evolution is put into $(u, v)$, the corresponding diagram to the left diagram in
figure \ref{fig:swing1} would describe a process where $(u, v)$ splits into $(u, z)$ and $(z, v)$,
both of which then scatter against $(x_1, y_1)$ and $(x_2, y_2)$. This is thus a multiple scattering
contribution, referred to as the ``fluctuation'' contribution in \cite{Iancu:2005nj}. 
If one carefully studies
all the possible Feynman graphs, then the following picture appears when dipoles in the
same cascade are allowed to interact. As mentioned above, first one of the dipoles $(x_1, y_1)$
and $(x_2, y_2)$ splits (say $(x_1, y_1)$) into two new dipoles. Then one of the two new dipoles
swing with $(x_2, y_2)$, and two newer dipoles are formed. Thus we have three new dipoles at
the end of the process. Any one of these three dipoles can then interact with $(u, v)$. The
$2 \to 3$ vertex which involves the swing goes like $\abar^3/N_c^2$, and is therefore colour suppressed.
There are also terms which cannot be interpreted in terms of dipole interactions. However,
these are all proportional to $\abar^3/N_c^4$ and they can therefore be neglected as compared to
the dipole swing contribution. It was shown in \cite{Avsar:2007xh} that the colour structures of all the
multiple scatterings diagrams exactly correspond to the colour structures of diagrams which
in the evolution of a dipole cascade can be interpreted in terms of the dipole swing and
the $k \to k+1$ vertices mentioned above. Whether or not one can thereby obtain a boost
invariant evolution is, however, not quite clear. A detailed study trying to attack this
problem, and to study the structure of the generated evolution equations is under way 
\cite{Avsar}. 

Boost invariance (sometimes referred to as the self-duality) of the evolution, and the
generalization of the B-JIMWLK hierarchy, have previously been discussed in a series of
papers \cite{Kovner:2005en, Iancu:2005nj, Hatta:2005rn, Levin:2005au}. 
Note, however, that the $k \to k+1$ vertices mentioned above give rise to
equations more general than the ``Pomeron Loop'' 
equations derived in \cite{Iancu:2005nj, Levin:2005au} 
(the toy model
analogy of this has been discussed for example in \cite{Iancu:2006jw}). 
As mentioned, it is, however, not
know whether or not one can obtain a boost invariant evolution in the full model. For
example, in \cite{Albacete:2006uv} it was shown that the higher order corrections arising from the strong
classical fields in the JIMWLK formalism contains quadrupoles, sextupoles and so on. It
may therefore be that one needs to include more complicated colour structures in a fully
consistent formalism.

In the MC implementation of our model \cite{Avsar:2006jy, Avsar:2007xg}, 
each dipole
is given one of $N_c^2$ possible colour indices, and only 
dipoles with the same colour index are allowed to swing\footnote{
The probability that a given colour--anti-colour 
pair forms a colour singlet is $1/N_c^2$.}. Here
the weight for the swing process was chosen so that it favours the formation of smaller
dipoles, which explicitely introduces saturation effects in the cascade evolution as smaller
dipoles both split and interact more weakly. As we discussed in \cite{Avsar:2006jy} 
this procedure also
approximates higher order multipoles, which might be needed in a consistent formulation
as mentioned above.
The 
similarity with the CGC formalism can then be explained 
as follows. In the CGC, it can be shown that the number 
density of gluons satisfy \cite{Iancu:2002xk}
\begin{eqnarray}
\frac{dN}{dY\,d^2\pmb{b}d^2\pmb{k}} \lesssim \frac{1}{\alpha_s}
\label{eq:CGClimit1}
\end{eqnarray}
due to saturation. If one integrates over rapidity, one rather 
gets 
\begin{eqnarray}
\frac{dN}{d^2\pmb{b}d^2\pmb{k}} \lesssim \frac{1}{\alpha_s^2}.
\label{eq:CGClimit2}
\end{eqnarray}
What happens in the CGC is that one can continue to pack 
the hadron with more gluons until there are so many gluons overlapping that 
their mutual interaction is strong enough to prevent further occupation at 
a particular $\pmb{b}$. In a semiclassical picture we can think of 
each gluon as a disc of radius $\sim 1/|\pmb{k}|$. Holding $\pmb{k}$ 
fixed, one will sooner 
or later reach a point where it is not possible to put in any more gluons 
of that $\pmb{k}$. In that case we have to increase $\pmb{k}$, which 
corresponds to adding smaller discs into the proton. At one stage 
those smaller discs will also fill up the avaliable holes, but there is then more
room for even smaller discs and so on. One can continue in this 
way forever,  with the typical gluon momenta being pushed to 
higher values, and 
the total number of gluons therefore never ceases to grow\footnote{The 
production rate of additional gluons does saturate however.}. 

The dipole swing works in a very similar way. 
Since the evolution is driven by the $1\to 2$ 
splitting 
plus the $2\to 2$ swing, the total number of dipoles will continue 
to grow forever. Assume, however, that we wish to put many dipoles of 
similar size $\pmb{r}$ around the same impact parameter
$\pmb{b}$. If the number of dipoles is less than $N_c^2$, there are no problems 
since the swing is not very likely. However, as soon as we have $N_c^2$ dipoles 
they can start to swing, and since in this case they almost sit on top of each other, 
they will do so as soon as the chance is given (the swing favours the 
formation of smaller dipoles, and in this case 
the swing probability is thus very large \cite{Avsar:2006jy}). 
When two dipoles swing they will 
be replaced by two smaller dipoles, with different impact parameters 
$\pmb{b}'$. This implies that the dipole occupation number satisfies
\begin{eqnarray}
\frac{dN}{d^2\pmb{b}d^2\pmb{r}} \lesssim N_c^2 \sim \frac{1}{\alpha_s^2}.
\label{eq:diplimit}
\end{eqnarray} 
Here we assume $\bar{\alpha}$ to be fixed and of order $1/\pi$, in which case 
we get $\alpha_s \sim 1/N_c$. When the number of smaller dipoles around $\pmb{b}'$ 
gets large enough they will in turn start to swing to produce even smaller dipoles and so on. 
Thus we get a picture which is similar to that in the CGC formalism. 

The dipole swing suppresses the growth of the cascade in the 
transverse plane, and it is interesting to see how large 
effects it has on the evolution. In the next section we will 
therefore present our results with and without the swing. 
Obviously, we cannot expect the swing to change the $Y$ dependence 
of $R_{bd}(Y)$ in a qualitative manner, in particular we cannot 
expect it to modify an exponential growth, since the associated 
interactions are still mediated by massless gluons.  


\section{Results}
\label{sec:res}

We will in this section present the results obtained 
from our MC simulation for the growth of $R_{bd}(Y)$. 
Let us first mention that it is extremely difficult 
to make analytic predictions as in section \ref{sec:bfkl}. 
In our simulations we take into account effects of 
energy-momentum conservation, which are related to, 
but go beyond, the next-to-leading order corrections  
to the cascade evolution. As mentioned in the 
previous section, we also include saturation effects in 
the evolution, and the full equations which include 
these effects are not known (even in the large $N_c$ limit), 
and once they are 
known they will probably be extremely complicated to solve. 
Furthermore, analytic estimates can never fully take 
into account the full impact parameter dependence 
of the evolution, which is relevant for the 
present study.  

The first results we show are for the case of a running 
coupling, and without any confinement effects. In order 
to avoid singularities, the value of $\alpha_s(p_\perp)$ 
is frozen below the scale $p_\perp = 2/r_{max}$ 
where $r_{max}$ is a free parameter which 
in our full model 
sets the confinement scale. In our previous studies 
\cite{Avsar:2007xg} we have set $r_{max}=$ 3.5 GeV$^{-1}$ 
which is also the value we will use in the present study.
($\Lambda_{QCD}$ is fixed to 0.22 GeV)

In the discussion in section \ref{sec:bfkl}, we considered 
the projectile to be an elementary dipole. Due to the 
fact that higher $Y$ values are extremely time consuming to 
simulate, we will here use a projectile which is more
dense initially. We will therefore consider $pp$ scattering
where the initial proton is modeled as consisting of
three dipoles (with a
Gaussian distribution in sizes determined by the scale $r_{max}$) 
in a triangular configuration as was discussed
in \cite{Avsar:2006jy, Avsar:2007xg}, where the frame independence of the process have also 
been demonstrated. We will therefore compute 
$T_Y(\pmb{b})$ in the CM frame, since this is numerically
the least time consuming frame. 

\FIGURE[t]{
  \includegraphics[angle=270, scale=0.55]{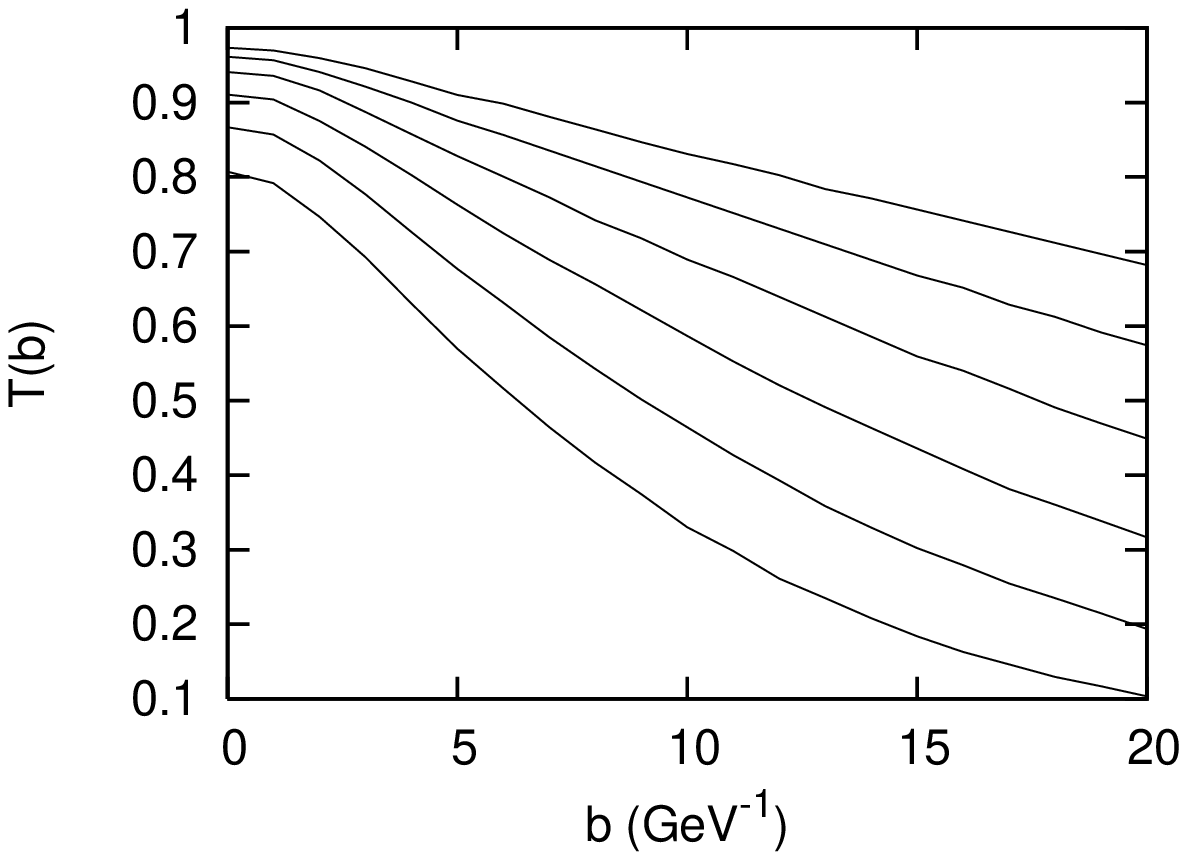}
  \includegraphics[angle=270, scale=0.55]{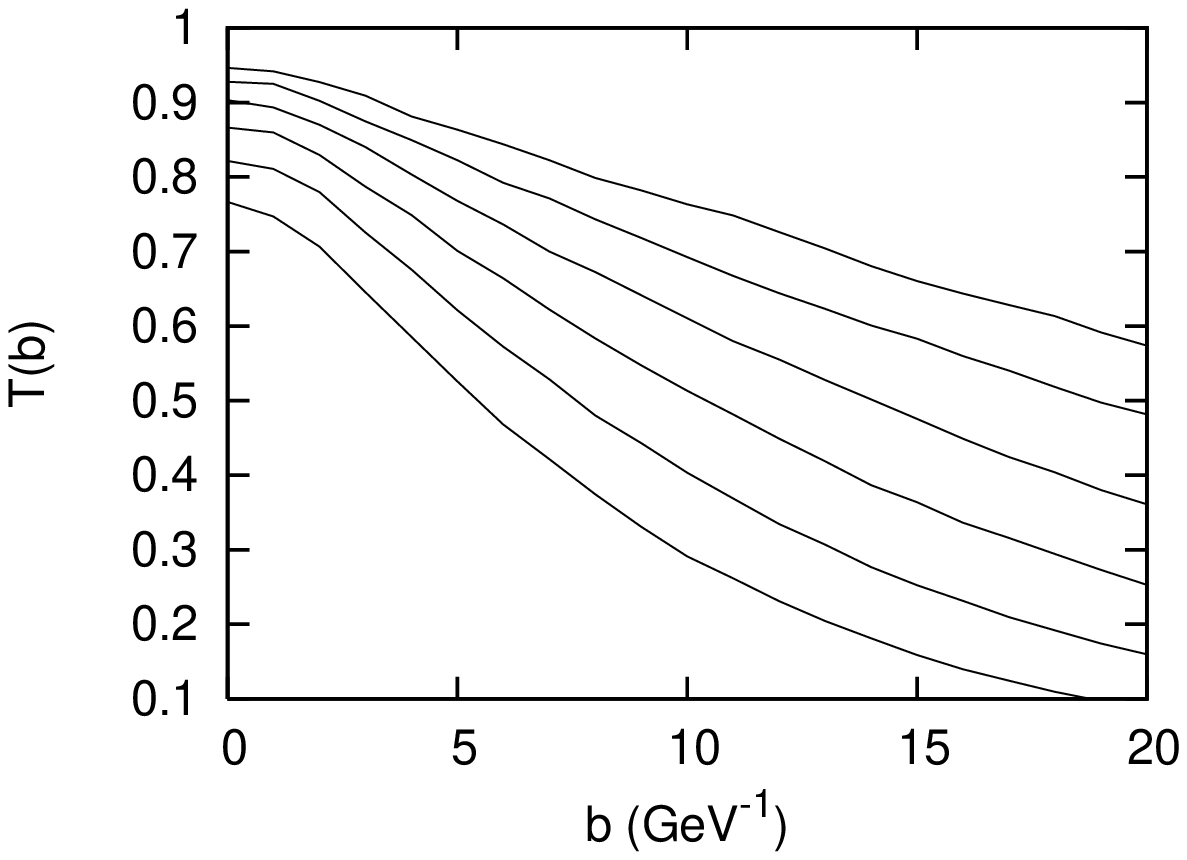}
  \caption{\label{fig:noconf}  The transverse profile of the scattering 
  amplitude $T$ including a running coupling, and without 
  confinement effects. The left figure excludes the 
  swing while the right figure includes it. In both figures 
  the lowest curve is calculated at $Y=8$, and $Y$ is increased by 
  2 units for each new curve. In both cases an exponential growth of 
  $R_{bd}(Y)$ is seen. The main mechanism driving the growth 
  is clearly the very fast growth of the white region. }}

\FIGURE[t]{
  \includegraphics[angle=270, scale=0.7]{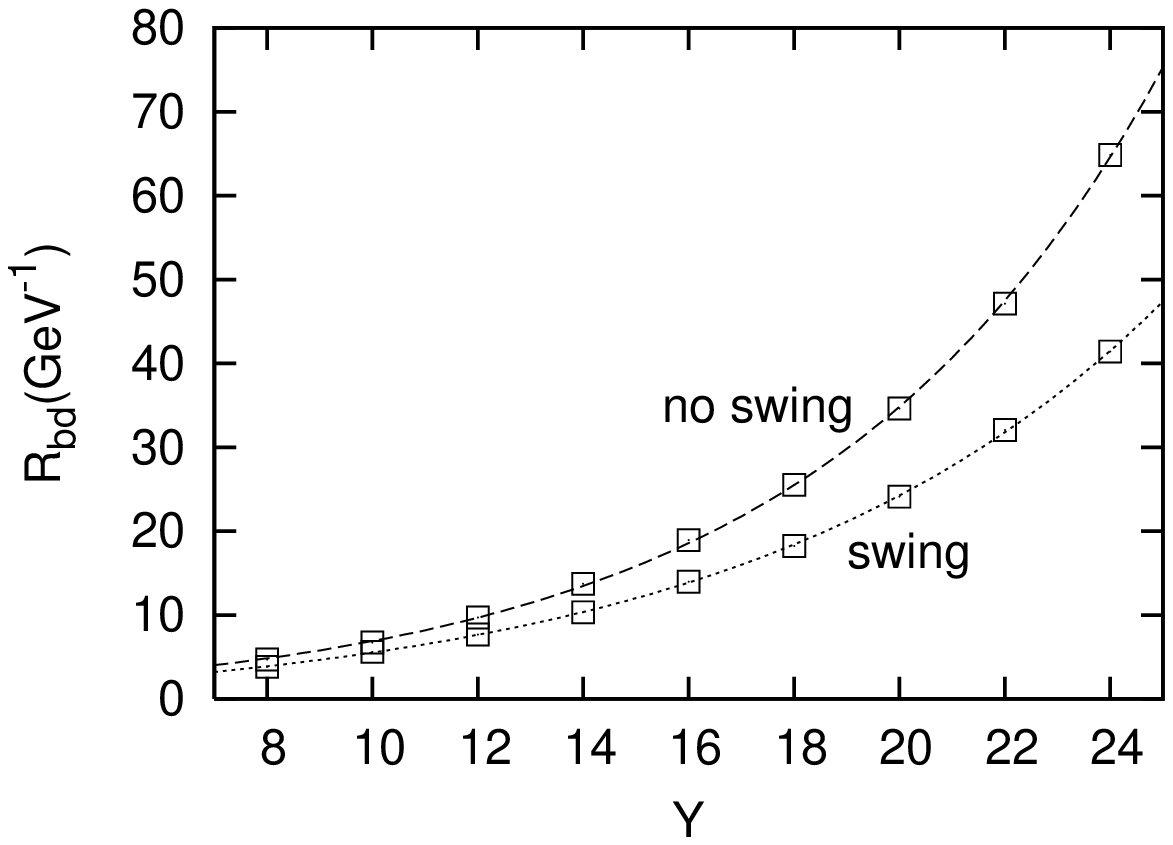} 
  \caption{\label{fig:Rbdnoconf}  The black disc radius $R_{bd}(Y)$ 
  plotted as a function of $Y$, as calculated from figure \ref{fig:noconf}. 
  The squares correspond to evolution including (lower set),
  and excluding (upper set) the swing. Together 
  with each curve, the corresponding fits are also shown. }}

Our first result are shown in figure \ref{fig:noconf}. Here we plot
$T_Y(b)$ as a function of $b$ (we average over the angle). 
We show two plots, including (right plot) and excluding (left plot) 
the swing. The lowest curves are calculated at $Y=8$, and $Y$ is 
increased by 2 units for each new curve. We see very 
large contributions from large impact parameters, and
the resulting profile is very flat. In this case $R_{bd}(Y)$ 
grows exponentially which can be seen in figure \ref{fig:Rbdnoconf}. 
Here, the upper curve excludes the swing, 
and we find that $R_{bd}(Y)$ can be fitted as $R_{bd}(Y) \approx 
$ 1.8$\cdot$exp$(0.15Y)$GeV$^{-1}$.
If the dipole swing is included we instead get $R_{bd}(Y)\approx 
2.0\cdot$exp$(0.13Y)$GeV$^{-1}$. Also in the case 
where we include the swing, we see that $R_{bd}(Y)$ grows 
very rapidly. As mentioned in the introduction this is partly 
due to the breakdown of perturbation theory since the value 
$\alpha_s(p_\perp)$ is very large during the evolution, which 
is moreover extremely sensitive to the infrared cutoff $r_{max}$. 

\FIGURE[t]{
  \includegraphics[angle=270, scale=0.7]{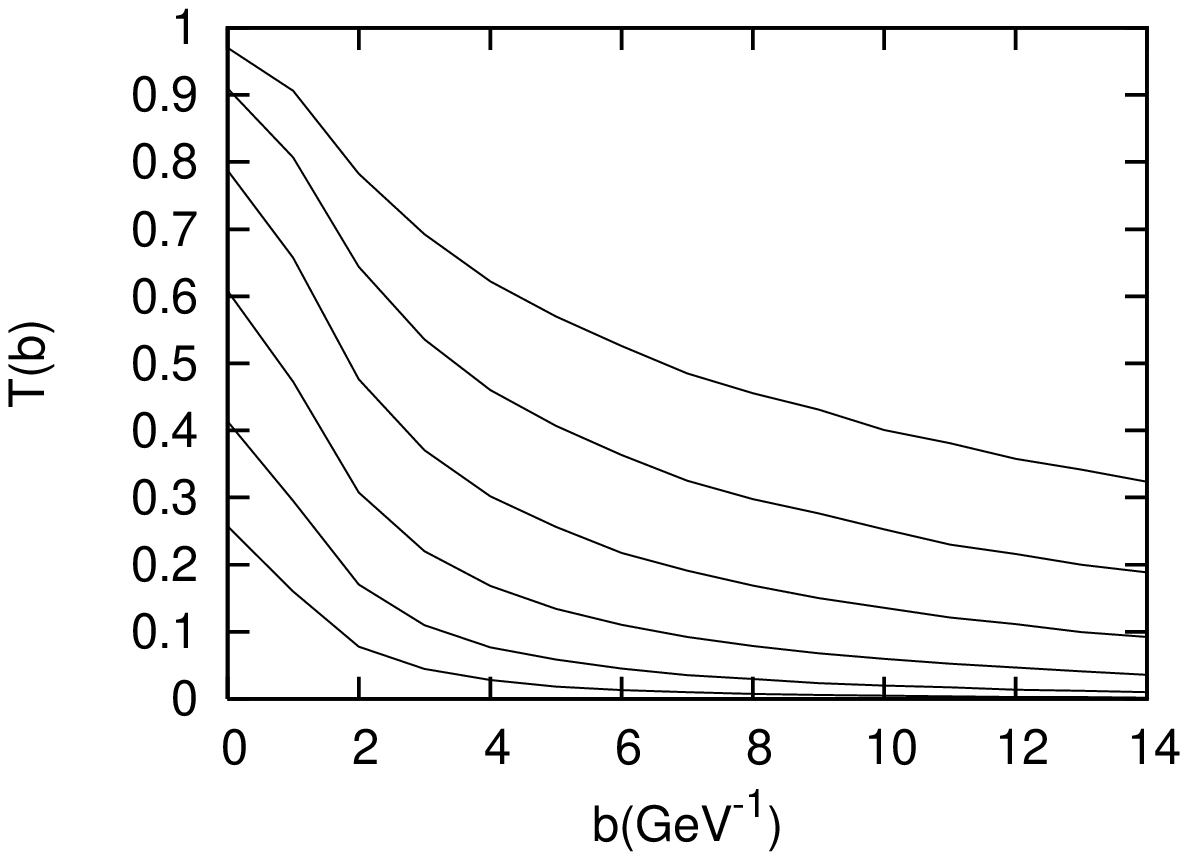} 
  \caption{\label{fig:onirunning}  The transverse profile of $T$ for 
    onium-onium scattering, without the swing. We again 
    find a very flat profile, with very large contributions 
    from large $b$. The lowest curve is calculated at $Y=12$, and 
    $Y$ is increased by 4 units for each new curve. }}

Of course, since in this case the initial dipoles are rather 
large (with sizes close to $r_{max}$) one is in the soft region already from 
the beginning. We therefore also study onium-onium
scattering, where the initial onia have small sizes, we 
take the two cases $r_0 = 0.5$ and $1$ GeV$^{-1}$ respectively 
(both initial onia have the same size). In figure 
\ref{fig:onirunning} we show the transverse profile for the case 
$r_0=1$GeV$^{-1}$. Here we do not include the swing and we 
once again find a very flat distribution implying an exponential
growth for $R_{bd}(Y)$. The result for the case $r_0=0.5$GeV$^{-1}$
looks essentially the same. 

Next we switch to a fixed coupling, $\alpha_s=0.2$. It is 
well known that the leading order dipole cascade also 
in this case shows a fast diffusion 
towards large dipole sizes. In \cite{Avsar:2005iz} we demonstrated
the very large effects of energy-momentum conservation 
on the evolution. In this case the production of both small (from 
$p_+$ conservation) and large (from $p_-$ conservation) dipoles
are suppressed, and the growth of the cross section is 
severely dampened. 

\FIGURE[t]{
  \includegraphics[angle=270, scale=0.7]{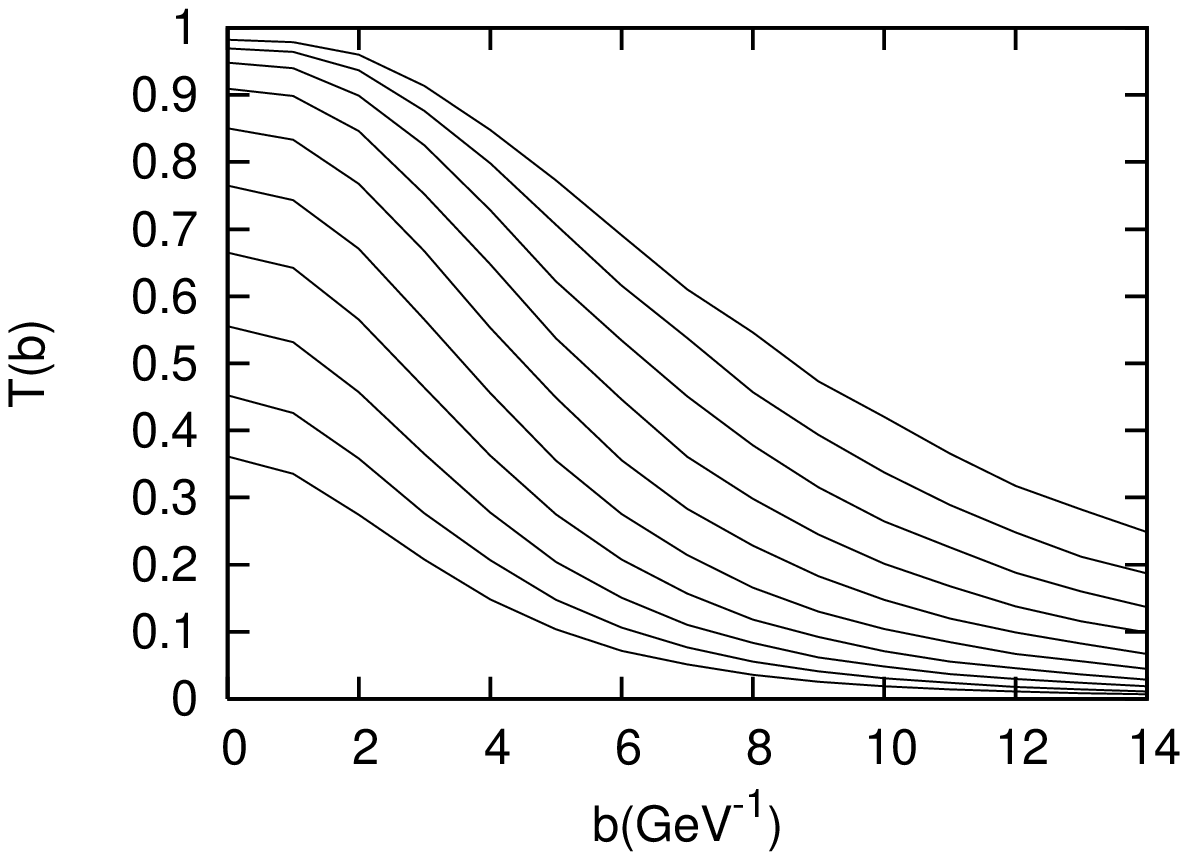}
  \caption{\label{fig:noconffixa}  The transverse profile of the 
  scattering amplitude for a fixed coupling, $\alpha_s=0.2$. 
  In this case the swing is not included. The lowest curve is 
  calculated at $Y=8$, and $Y$ is increased by 2 units for 
  each successive curve. Although 
  the growth of $R_{bd}(Y)$ seems to be linear for lower $Y$, we 
  can see that the white region grows very fast, implying an 
  increasingly faster growth of $R_{bd}(Y)$ as $Y$ increases.} }

\FIGURE[t]{
  \includegraphics[angle=270, scale=0.7]{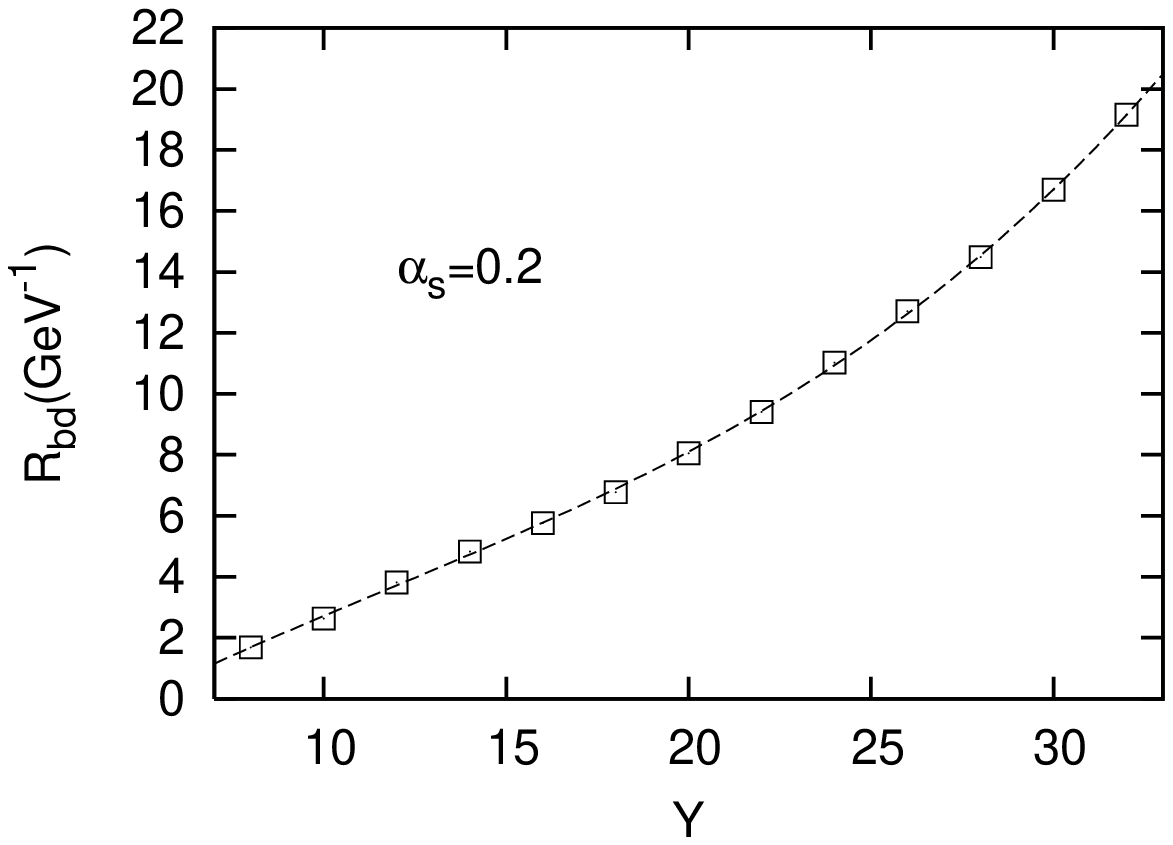} 
  \caption{\label{fig:Rbdnoconffixa}  The black disc radius $R_{bd}(Y)$
    as a function of $Y$ as calculated from figure \ref{fig:noconffixa} 
    (squares).
    In this case $R_{bd}(Y)$ can be fitted by a polynomial (the dashed
    line) as explained in the text. }}

The transverse profile of the scattering amplitude is shown in figure
\ref{fig:noconffixa}. As compared to figure \ref{fig:noconf} 
we see that the growth is significantly reduced. This is 
because the unrealistically large contributions from larger $b$
present in the running coupling case are much reduced. 
From figure \ref{fig:noconffixa} it seems that, 
for rapidities under $Y \approx 20$, the growth is even 
linear\footnote{In order to see the FM bound-breaking growth sooner, 
we have here defined $R_{bd}$ for $a= 0.3$ in \eqref{eq:blackdiscdef}. 
This will of course affect the 
absolute value of $R_{bd}$, but it does not change the fact that $R_{bd}$ grows more 
than linearly with $Y$.}. However, as $Y$ increases, we clearly see a deviation
from the linear growth, which becomes greater for the largest $Y$. 
Clearly, the FM bound-breaking growth arises not because the 
central regions reach the black disc limit too fast, but 
rather because the white region, where the scattering is very 
weak initially, turns grey too fast. This indeed confirms 
the expectation that the FM bound is in this case violated 
due to long range contributions coming from the perturbative Coulomb 
fields \cite{Kovner:2002yt}. What is interesting, however, is the 
fact that the growth of $R_{bd}(Y)$ is much slower than 
expected from the leading order evolution. This is clearly 
illustrated in figure \ref{fig:Rbdnoconffixa}. What we see 
here is that the 
shape of $R_{bd}(Y)$ cannot be fitted by an exponential, 
at least for $Y$ up to 32. The growth accelerates as $Y$ 
increases, so as $Y\to \infty$ the growth should eventually 
reach an exponential (indeed the white region already grows
exponentially). For the present values we instead 
find that the shape can be fitted by a polynomial, 
and in figure \ref{fig:Rbdnoconffixa} we show a fit 
$R = (-3.3 + 0.76\cdot Y - 0.02\cdot Y^2 + 6\cdot 10^{-4}\cdot Y^3)$GeV$^{-1}$, 
which is the best fit we have found. Such a fit would 
imply a cross section growing like $\sigma_{tot} \sim $ln$^6(s/s_0)$. 
This growth is further reduced if we include the dipole swing, as we 
illustrate in figures \ref{fig:noconffixa2} and \ref{fig:Rbdnoconffixa2}. 
Even if the growth seems to be linear, one can again see that 
the white region expands rapidly and the growth of $R_{bd}$ accelerates 
as $Y$ increases. The fit in figure \ref{fig:Rbdnoconffixa2} corresponds 
to a quadratic fit.

\FIGURE[t]{
  \includegraphics[angle=270, scale=0.7]{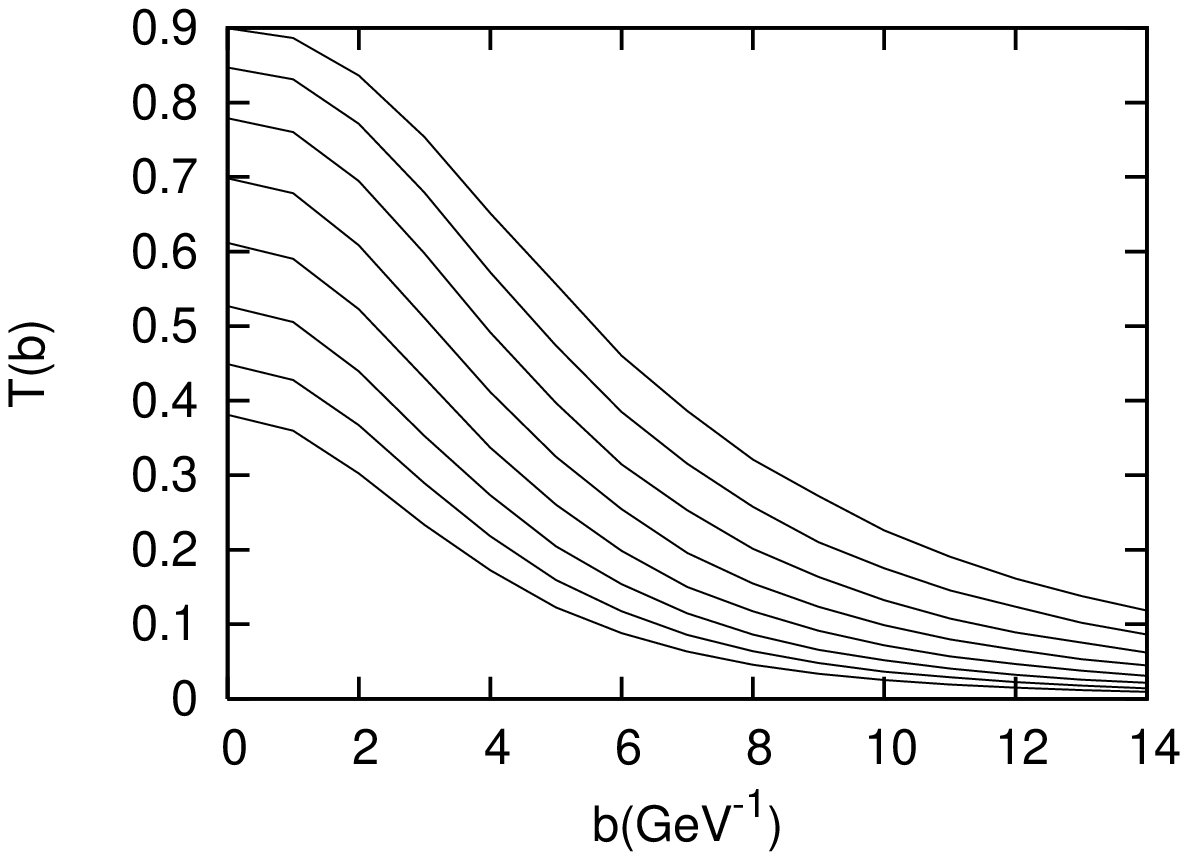} 
  \caption{\label{fig:noconffixa2}  The transverse profile for fixed coupling 
    and including the swing. The lowest curve is calculated at $Y=10$ and 
    $Y$ is increased 2 units for each new curve. }}

\FIGURE[t]{
  \includegraphics[angle=270, scale=0.7]{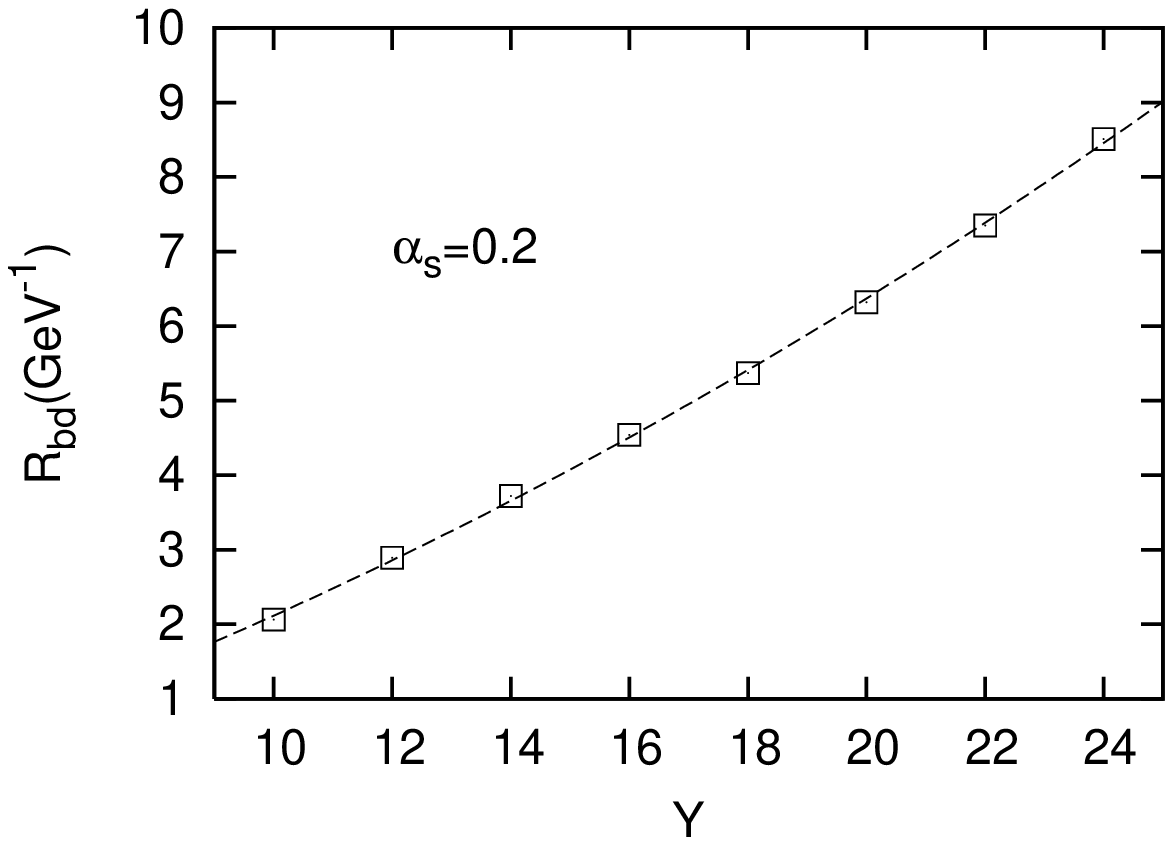} 
  \caption{\label{fig:Rbdnoconffixa2}  The black disc radius $R_{bd}(Y)$
    as a function of $Y$ as calculated from figure \ref{fig:noconffixa2} 
     }}

In order to satisfy the FM bound we must introduce 
a scale above which the gluon fields falls off exponentially. 
In \cite{Avsar:2007xg} we replaced the 
momentum space Coulomb propagators, $1/\pmb{k}^2$, by 
screened propagators, $1/(\pmb{k}^2+M^2)$, where $M=1/r_{max}$.
The dipole splitting kernel, for the process $(\pmb{x},\pmb{y})
\to (\pmb{x},\pmb{z}) + (\pmb{z},\pmb{y})$, is then modified as 
\begin{eqnarray}
\frac{(\pmb{x} - \pmb{y})^2}{(\pmb{x} - \pmb{z})^2
(\pmb{z} - \pmb{y})^2} \to \biggl ( \frac{1}{r_{max}}
\frac{\pmb{x} - \pmb{z}}{|\pmb{x} - \pmb{z}|}K_1(
|\pmb{x} - \pmb{z}|/r_{max}) - \frac{1}{r_{max}}
\frac{\pmb{z} - \pmb{y}}{|\pmb{z} - \pmb{y}|}K_1(
|\pmb{z} - \pmb{y}|/r_{max}) \biggr )^2. \nonumber \\
\label{eq:dipkernel}
\end{eqnarray}
Similarly the dipole--dipole scattering amplitude $T_0(\pmb{x},
\pmb{y}|\pmb{u},\pmb{v})$ 
is modified as 
\begin{eqnarray}
\frac{\alpha_s^2}{2}\mathrm{ln}^2\biggl \{
\frac{|\pmb{x} - \pmb{v}||\pmb{y} - \pmb{u}|}
{|\pmb{x} - \pmb{u}||\pmb{y} - \pmb{v}|} \biggr \}\to
\frac{\alpha_s^2}{2} \biggl ( K_0(|\pmb{x} - \pmb{u}|/r_{max})
 - K_0(|\pmb{x} - \pmb{v}|/r_{max}) - \biggr . \nonumber \\ 
\biggl . K_0(|\pmb{y} - \pmb{u}|/r_{max}) 
+ K_0(|\pmb{y} - \pmb{v}|/r_{max}) \biggr )^2.
\label{eq:dipamp}
\end{eqnarray}
Here $K_0$ and $K_1$ are modified Bessel functions which 
both behave as $K(x) \sim \sqrt{\frac{2}{x}}e^{-x}$ for
large $x$. In the analysis of section \ref{sec:bfkl}, this 
would imply an exponentially decaying profile in $\pmb{b}$,
which would compensate the exponential growth in $Y$ 
of the BFKL solution. The evolution should then 
satisfy the FM bound. This has been used in \cite{Ferreiro:2002kv} 
to estimate the constant $C$ in \eqref{eq:FM} as
\begin{equation}
C = 2\pi \biggl ( \frac{\omega}{\mu}\biggr )^2,
\label{eq:edmondcoff}
\end{equation}
where $\omega$ was defined in \eqref{eq:Tbfkl}, and 
$\mu$ is the confinement scale, \emph{i.e.} $r_{max}^{-1}$ 
in our model. The constant $C$ can thus roughly be determined 
by combining the hard pomeron intercept with the
non-perturbative confinement scale, albeit in a heuristic 
fashion. 

Recently it has been shown that experimental results 
on $pp$ collisions favours a ln$^2 s$ fit (rather than a 
ln$s$ fit) 
to the total $pp$ cross section \cite{Block:2005pt}. 
The fit in \cite{Block:2005pt} has the form 
\begin{equation}
\sigma \approx c_0 + c_1\,\mathrm{ln}\biggl (\frac{s}{2m^2}
\biggr) + c_2\,\mathrm{ln}^2\biggl (\frac{s}{2m^2}
\biggr),
\label{eq:BHfit}
\end{equation}
where we have neglected terms which fall off as a power 
of $s$. The various coefficients above were found to be 
$c_0 \approx 37$ mb, $c_1 \approx -1.4$ mb and $c_2 \approx 0.28$ mb. 
In \cite{Ferreiro:2002kv}, the value of $\mu$ has been argued to be around
2$m_\pi\approx 0.28$ GeV, which interestingly is  
equal to $r_{max}^{-1}$ with $r_{max}=3.5$ GeV$^{-1}$. 
In this case the value of \eqref{eq:edmondcoff}
would be around 9 mb, considerably higher than $c_2$ above. 
Of course, the value of \eqref{eq:edmondcoff} is not supposed
to reproduce $c_2$ since it has been derived under rather 
crude assumptions (any attempt to include nonperturbative effects 
will admittedly be heuristic as well). It is also a well known fact that the leading order 
BFKL exponent $\omega$ is too large to fit data. If we would 
for example replace $\omega$ by its NLO value, $\omega \approx 0.3$,
$C$ would be reduced almost by a factor of 4. 

\FIGURE[t]{
  \includegraphics[angle=270, scale=0.55]{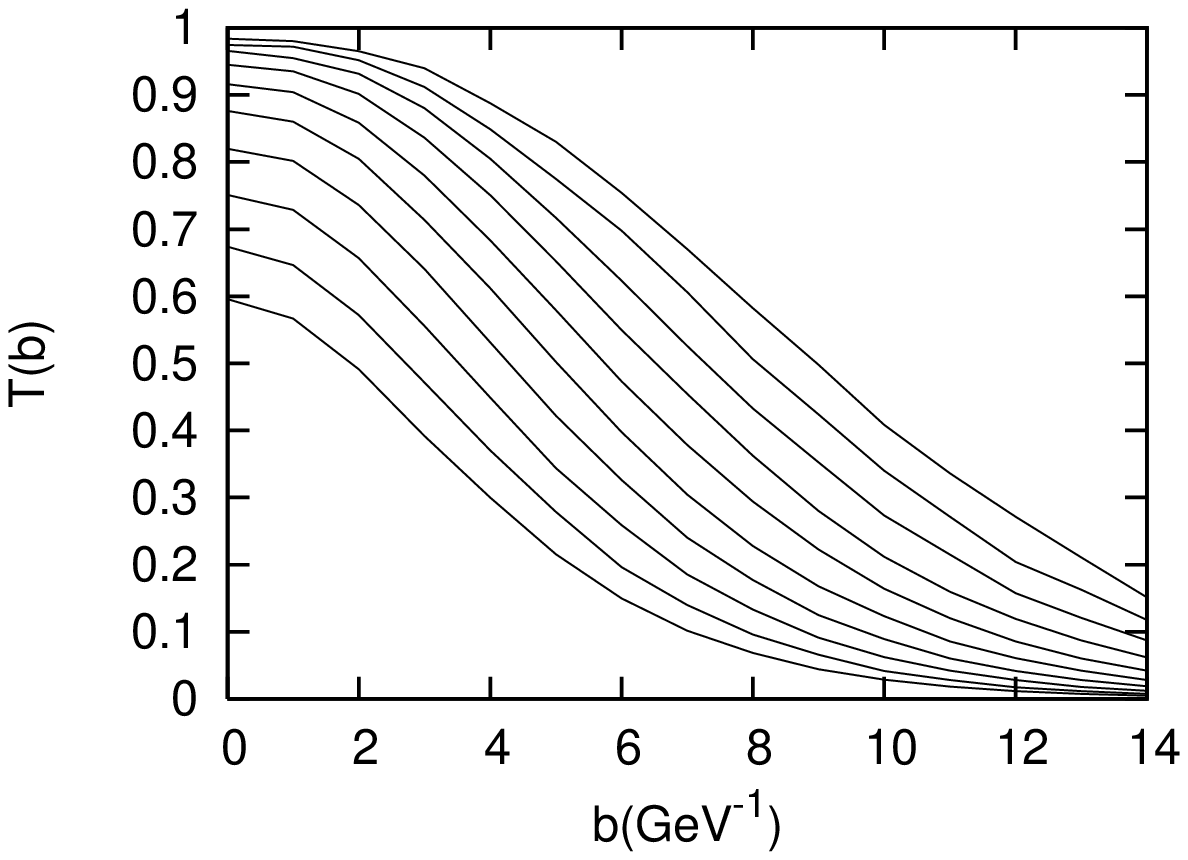}
  \includegraphics[angle=270, scale=0.55]{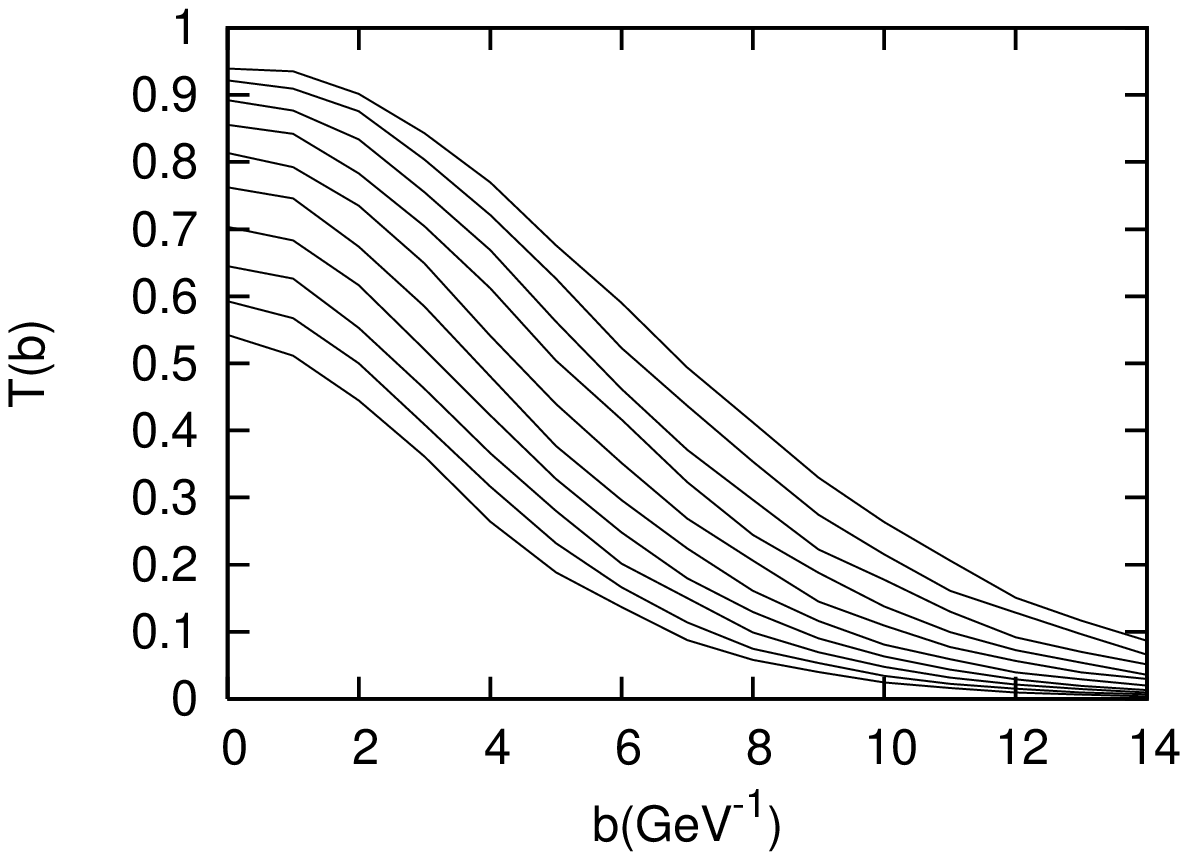}
  \caption{\label{fig:conf} The transverse profile of $T$ 
    including confinement effects via equations \eqref{eq:dipkernel} 
    and \eqref{eq:dipamp}. In the left figure the swing is 
    excluded while it is included in the right figure. 
    The lowest curves are calculated at $Y=8$, and $Y$ is increased 
    by 2 units for each new curve. Here the growth is clearly linear, 
    and as expected we see that the growth of the white region is considerably
    suppressed. }}

In figure \ref{fig:conf} we show the transverse profile of the 
scattering amplitude, calculated using \eqref{eq:dipkernel} and 
\eqref{eq:dipamp}, and using a running coupling. 
Note that, as compared to figure \ref{fig:noconffixa}, 
the difference now is that the growth of the white region is 
considerably suppressed, as expected from confinement. 
In this case 
we can see a linear growth, which is illustrated in figure 
\ref{fig:Rbdconf}. 
Here we find that $R_{bd}$ can be fitted as 
$R_{bd}(Y) = (-1.36 + 0.39Y)$GeV$^{-1}$ when the swing is excluded, 
and as $R_{bd}(Y) = (-1.70 + 0.31Y)$GeV$^{-1}$ when it is included. 
This would imply a 
cross section growing as $\sigma \sim 2\pi \cdot 0.39^2$ ln$^2 s$ 
GeV$^2 = 0.37\cdot$ln$^2 s$ mb for the former case, while for 
the latter case we would have $\sigma \sim 2\pi \cdot 0.31^2$ ln$^2 s$ 
GeV$^2 = 0.24\cdot$ln$^2 s$ mb.
The results are indeed quite close to 
the $c_2$ term in \eqref{eq:BHfit}. We might also ask how good the  
approximation $\sigma \sim 2\pi R_{bd}^2(Y)$ is. 
\FIGURE[t]{
  \includegraphics[angle=270, scale=0.7]{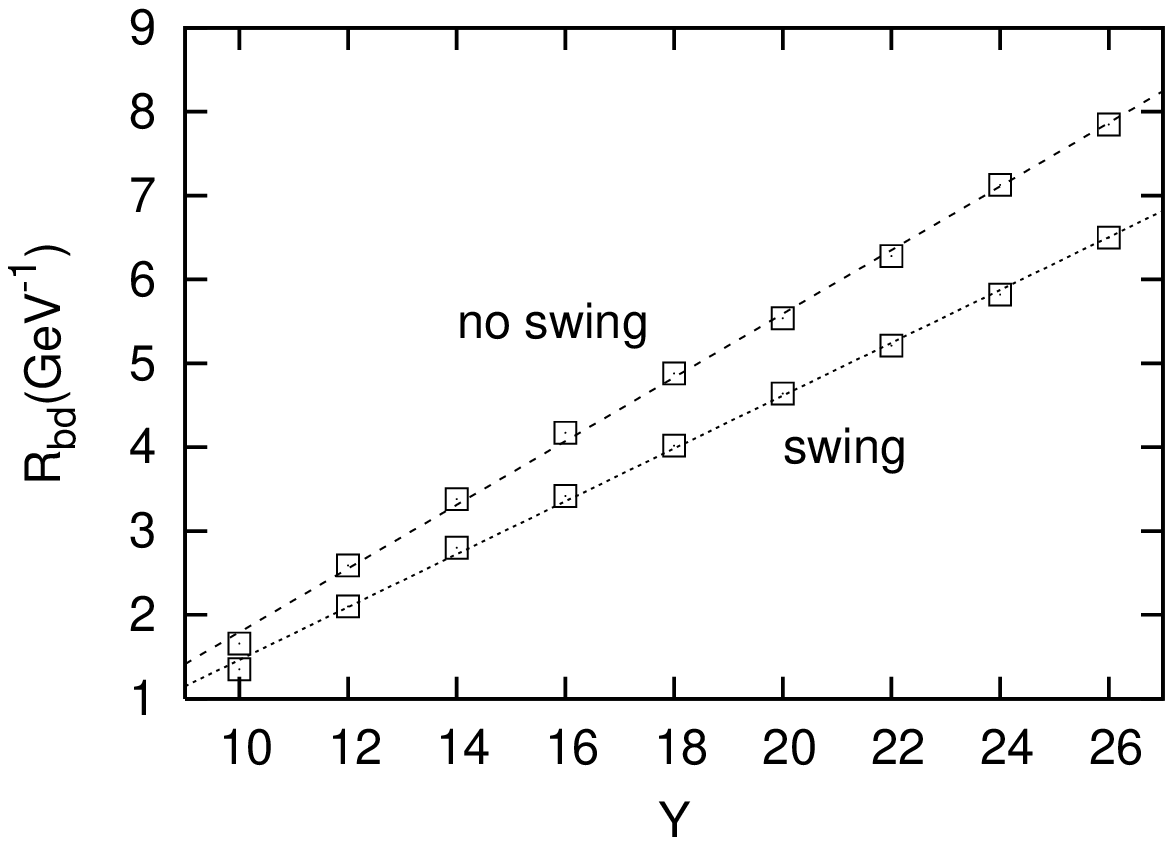}
  \caption{\label{fig:Rbdconf} The black disc radius $R_{bd}(Y)$ 
    calculated from figure \ref{fig:conf} excluding (upper set 
    of squares) and including (lower set of squares) the swing.
    Lines represent linear fits to the squares. } }
The MC results for the transverse profiles in figure \ref{fig:conf} 
can be estimated rather well (for fixed $Y$ ) by simple Gaussians,
$T=T(0)$ exp$(-c\cdot b^2)$. Thus, approximating these curves by such functions\footnote{
Given the fact the functions $K_1$ in the kernel \eqref{eq:dipkernel} 
fall exponentially one might think that an exponential
function would better approximate the $b$-profile. The good Gaussian 
approximation is related to our initial
proton model which has a Gaussian distribution \cite{Avsar:2007xg}. 
Actually our $b$-profile has a somewhat longer tail than
a Gaussian for large $b$ due to the fluctuations in the evolution, as was discussed in 
\cite{Avsar:2007xg}.} the cross section would be given by 
\begin{equation}
\sigma_1 = 2 \int d^2\pmb{b}\, T(0)\, \mathrm{exp}(-c\cdot b^2)
 = \frac{\pi}{c}T(0).
\end{equation}
On the other hand, $R_{bd}$ is defined as $T(R_{bd})= a$ 
with $a = \mathcal{O}(0.5)$, and the approximation 
$\sigma \sim 2\pi R_{bd}^2(Y)$ then gives 
\begin{equation}
\sigma_2 = \frac{2\pi}{c}\,\mathrm{ln}\biggl (\frac{T(0)}{a}
\biggr ).
\end{equation}
Thus 
\begin{equation}
\frac{\sigma_2}{\sigma_1} = \frac{2}{T(0)}\,\mathrm{ln}\biggl (\frac{T(0)}{a}
\biggr ). 
\end{equation}
For the Tevatron for example, we have $T(0) \approx 0.7$, 
while in figure \ref{fig:Rbdconf} we have set $a \approx 0.6$. This 
would give $\sigma_2/\sigma_1 \approx 0.5$. For higher 
energies where $T(0) \to 1$ we get $\sigma_2/\sigma_1 \approx 1$ 
so that the approximation $\sigma \sim 2\pi R_{bd}^2(Y)$ works
reasonably well. 

Of course we might as well directly calculate the total cross section
using the MC, such as we did in \cite{Avsar:2007xg}.
Thus we can try to fit a curve of the form \eqref{eq:BHfit} 
to our results \cite{Avsar:2007xg} for the total $pp$ cross section. 
We thus parameterize the cross section as
\begin{equation}
\sigma = A + B \,\mathrm{ln} s + C\, \mathrm{ln}^2 s,
\label{eq:ourfit}
\end{equation}
where $s$ is measured in units of GeV$^2$. In this case 
we find the results $A = 34$ mb, $B = -1.8$ mb 
and $C = 0.30$ mb which are quite close to the values 
$c_0$, $c_1$ and $c_2$ in \eqref{eq:BHfit}. We also try 
a fit $\sigma = D + E$ ln$^2s$ which works equally well, and 
we find $D = 20$ mb and $E= 0.24$ mb. We thus find that 
the Froissart bound is saturated, and 
by combining the 
perturbative evolution with nonperturbative confinement 
effects we moreover obtain a value for $C$ which is consistent with 
data. 

\section{Conclusions}
\label{sec:conc}

We have in this paper studied the growth of the black disc 
radius $R_{bd}(Y)$ for hadronic collisions, using our 
model developed in \cite{Avsar:2005iz, Avsar:2006jy, Avsar:2007xg}. 
Using a purely perturbative 
approach one cannot expect the FM bound to be satisfied, 
and for the running coupling case we indeed find an 
exponential growth for $R_{bd}(Y)$. On the other hand when
a fixed coupling is used we again find a fast growth of
$R_{bd}(Y)$, but the growth is in this case not an 
exponential as in the running coupling case, but can 
be fitted by a polynomial (at least for rapidities 
up to 32 units, corresponding to $s\sim 10^{14}$GeV$^2$). 
This is due to our inclusion
of energy-momentum conservation effects in the evolution, 
which severely dampens the leading order growth. However, 
it can be seen that the white region expands exponentially, 
which should imply an exponential growth for $R_{bd}(Y)$ 
eventually. The 
fact that the running coupling case shows such a fast 
growth is because, without any suppression 
of large dipoles, the coupling gets very large during
the evolution. Moreover, one gets unrealistically 
large contributions from large transverse separations.  

We model confinement effects by replacing the Coulomb 
propagators by screened propagators, in which case 
$R_{bd}(Y)$ grows linearly with $Y$, implying that the FM
bound is actually saturated. Furthermore, including
saturation effects during the cascade evolution, we see that we
obtain a value for the coefficient $C$ in \eqref{eq:FM} which 
is consistent with data. 

Obviously, our specific model used here makes sense only 
when including nonperturbative effects since the initial dipoles 
in the proton are quite large. However, also in the case 
where we start from a, smaller, single dipole did we see 
a very fast growth. It is also interesting to see what would
happen if one starts with a system containing perturbative dipoles, 
which is at the same time quite dense. Then we would expect
saturation to slow down the evolution, although we will of course
still get an exponential growth if no confinement effects 
are included. An interesting initial model is a proton consisting 
of three ``hot spots'', \emph{i.e.}
three saturated spots inside the proton. This model was discussed 
in \cite{Kovner:2002xa}. The sizes of these spots, related to the scale of chiral 
symmetry breaking, is estimated to be around 0.3 fm. Thus 
one may at least initially neglect confinement effects for the 
evolution of each spot.

To test the sensitivity of our model to the initial 
assumptions, we have also tried a model where 
the proton initially consists of 6 dipoles
in 3 spots (2 dipoles in each spot), where each spot has 
a size around $0.3$ fm. Including the swing, and 
with a running coupling and no confinement effects, 
we again find a very fast growth of $\sigma_{tot}$. For spot sizes 
of around $0.3$ fm, we find that the cross section 
shows a $s^{0.21}$ behaviour. 
 
Finally, when confinement is included we
find that a fit $D+ E\cdot$ln$^2s$ gives, $E\approx 0.31$mb 
which is higher than the result 0.24 mb found above. However, 
we should mention that  this initial 
model also reproduces the avaliable high energy data rather well. 
Furthermore we also get good results, as in \cite{Avsar:2007xg}, 
for the diffractive and elastic cross sections, both in 
$pp$ collisions and in DIS. To the accuracy of our model, 
we therefore get a reasonable description 
of data also with such an initial model.   

\section*{Acknowledgments}

I am grateful to G\"osta Gustafson for his encouragement
and support when this work was initiated in Lund. I would 
also like to thank Edmond Iancu for his critical remarks 
and for helpful discussions.

\bibliographystyle{utcaps}
\bibliography{refs}

\end{document}